\documentclass[journal]{IEEEtran}
\usepackage{graphicx}
\usepackage{amssymb}
\usepackage{color}
\begin{document}
\title{Suppression of Surface-Originated Gate-Lag by a Dual-Channel AlN/GaN HEMT Architecture}
\author{David A. Deen, \emph{Member, IEEE}, David F. Storm, D. Scott Katzer, \emph{Senior Member, IEEE}, R. Bass, David J. Meyer, \emph{Senior Member, IEEE}}
\thanks{D. A. Deen was with the Naval Research Laboratory, Electronic Science and Technology Division, SW Washington, DC 20375 USA. He is now with Seagate Technology, Read Head Operations, Bloomington,  MN 55435,  (e-mail: david.deen@alumni.nd.edu.)}
\thanks{D. F. Storm, D. S. Katzer, and D. J. Meyer are with the Naval Research Laboratory, Electronic Science and Technology Division, SW Washington, DC 20375 USA.}%
\thanks{R. Bass is with Sotera Defense Solutions, Herndon VA 20171-5393.}
\maketitle
\begin{abstract}
A dual-channel AlN/GaN high electron mobility transistor (HEMT) architecture is demonstrated that leverages ultra-thin epitaxial layers to suppress surface-state related gate lag. Two high-density two-dimensional electron gas (2DEG) channels are utilized in an AlN/GaN/AlN/GaN heterostructure wherein the top 2DEG serves as a quasi-equipotential that screens potential fluctuations resulting from surface and interface trapped charge. The bottom channel serves as the transistor's modulated channel. Dual-channel AlN/GaN heterostructures were grown by molecular beam epitaxy on free-standing HVPE GaN substrates where 300 nm long recessed and non-recessed gate HEMTs were fabricated. The recessed-gate HEMT demonstrated a gate lag ratio (GLR) of 0.88 with no collapse in drain current and supporting small signal metrics $f_t/f_{max}$ of 27/46 GHz. These performance results are contrasted with the non-recessed gate dual-channel HEMT with a GLR of 0.74 and 82 mA/mm current collapse with $f_t/f_{max}$ of 48/60 GHz.
\end{abstract}
\IEEEpeerreviewmaketitle
\IEEEPARstart{P}{erformance} degradation modes in GaN-based high electron mobility transistors (HEMTs) have received extensive scrutiny since the device's inception. Such modes include dc current collapse, dc-RF frequency dispersion due to the virtual-gate effect, gate and drain lag, and power slump. Many of these performance impairments have been traced back to surface and bulk trapping \cite{Binari},\cite{Wang}. Post-growth passivation techniques have become the most popular method to address the deleterious effects of surface state traps and include conformal oxide and nitride depositions \cite{Medjdoub},\cite{Lee},\cite{Huang},\cite{Koehler}. Surface chemical treatments have been investigated to minimize the effects of virtual gating on frequency performance \cite{Meyer},\cite{Wang2},\cite{Liu}. Alternative approaches for epitaxial passivation have also been given some attention \cite{Shen}. In an attempt to ascertain the origin of surface states in GaN-based HEMTs, Higashiwaki et al. reported on the formation of surface states in AlGaN barriers due to the high temperature anneal process that alloys the metallic ohmic contacts \cite{Higashiwaki}. Based upon this premise, reports of regrown ohmic contacts have shown that by avoiding the high temperature anneal process, the surface state density is reduced \cite{Xing1}. Despite nearly two decades of ongoing work in these areas, HEMT performance still has yet to reach many of the theoretical limits. This suggests there are more performance gains to be obtained by the technology if the sources of these modes can be addressed, either by materials or device design.

\begin{figure}
\includegraphics[width=0.99\columnwidth]{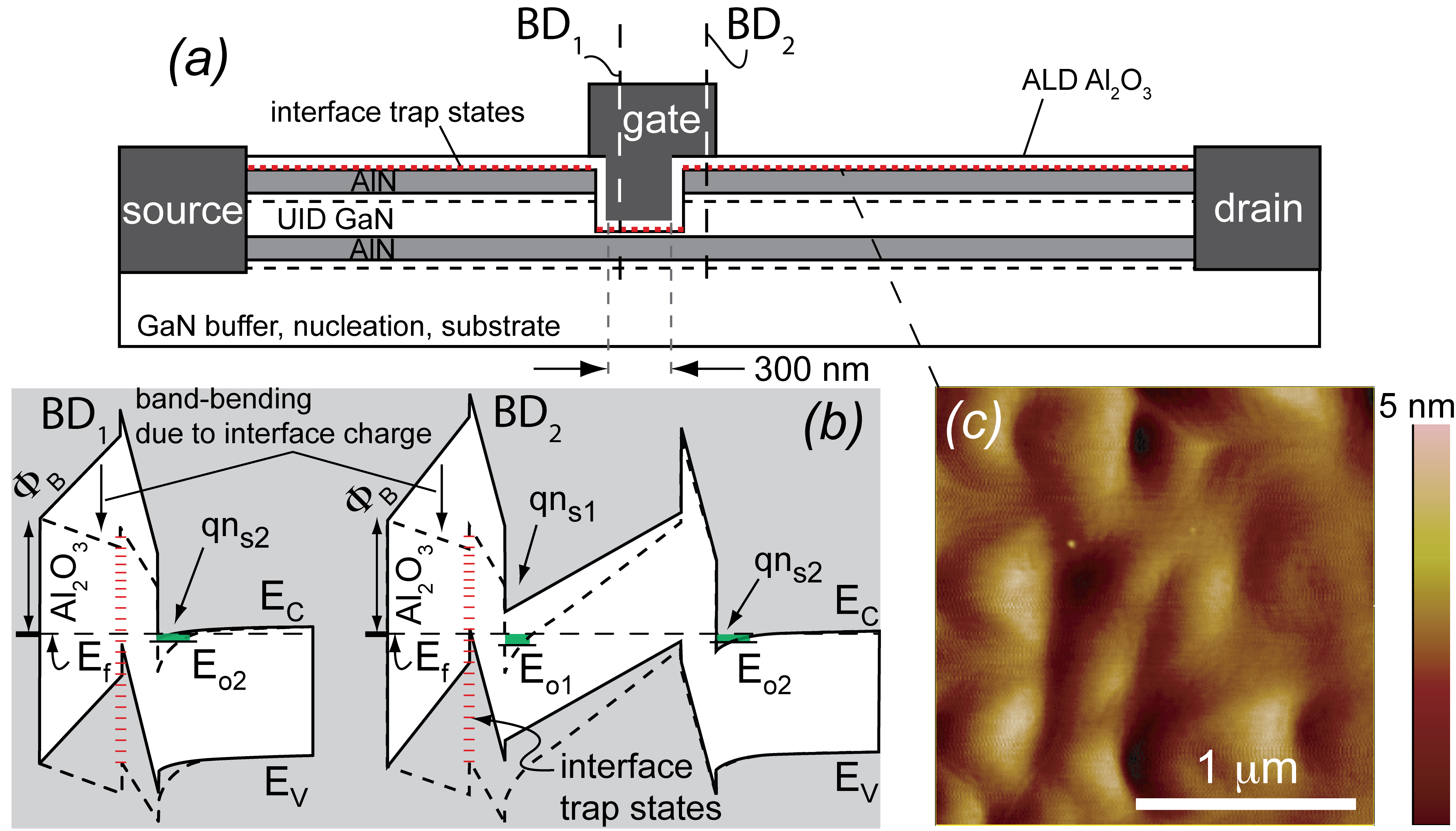}
\caption{(a) Cross-sectional illustration of the recessed-gate HEMT with corresponding band diagrams (BD) taken vertically through the gated regions. BD$_1$ corresponds to the gated intrinsic region of the HEMT with only a single channel and BD$_2$ corresponds to the access region of the HEMT with two coincident channels. Band diagrams are shown including (dashed) and excluding (solid) 6$\times$10$^{13}$ cm$^{-2}$ trapped charge at the oxide/AlN interface. Interfacial trap states are depicted by red dashes in the device cross-section and band diagrams. (c) Surface morphology of the as-grown AlN surface with a RMS roughness of 0.64 nm is shown by a $2\times2$ $\mu m^2$ AFM scan.}
\label{fig:xsection}
\end{figure}
The binary-barrier AlN/GaN HEMT has set remarkable performance benchmarks due to the exceptionally high polarization-induced 2DEG density achievable (up to 6$\times$10$^{13}$ cm$^{-2}$) with high mobility (1800 cm$^2$/Vs) \cite{Medjdoub,Zimmermann,Shinohara,Deen2}. Yet only a handful of HEMT designs have leveraged a few of the attributes that are inherent to this particular heterostructure \cite{Shinohara,Cao3,Deen3,Ganguly}. In this letter we propose and demonstrate a novel alternative to post-growth surface passivation based on a dual-channel AlN/GaN/AlN/GaN heterostructure. The upper AlN/GaN heterojunction undergoes a recess etch, conformal oxidation, and gate metal deposition as illustrated in Fig. \ref{fig:xsection}. The upper polarization-doped 2DEG serves solely to screen the potential fluctuations generated by surface trapped charge that would otherwise impose channel depletion leading to current collapse and gate lag. The trapped charge can also act as a source of remote ionized impurities that can scatter mobile channel electrons leading to mobility reduction in the current-carrying channel \cite{Deen1}. The bottom 2DEG serves as the gate-modulated channel. The HEMT access region includes both channels. Therefore, purely dual-channel AlN/GaN/AlN/GaN HEMTs have also been fabricated on the same wafer, serving as both a calibration structure for $CV$ and $IV$ characterization as well as a proxy to the recessed-gate HEMT access region. Several reports have been made on nitride-based dual-channel HEMTs with AlGaN or AlInN barriers with the intent to increase drain current density or assess HEMT noise characteristics and subsequently disregarded gate lag performance \cite{Chu},\cite{Jha},\cite{Zhang}. A notable attribute of using the AlN/GaN heterostructure for the HEMT design reported in this work is that the AlN barrier layers are inherently thin ($<$ 5 nm), which allows extremely shallow channels and therefore, multiple channel designs to maintain channels in close proximity to the surface. This is not the case for alloyed barriers that require a sufficient thickness in order to induce 2DEG formation.

AlN/GaN heterostructures were grown by RF-plasma assisted molecular beam epitaxy (MBE) on free-standing hydride vapor phase epitaxy (HVPE) GaN substrates. All heterostructures were grown at 650$^o$ C and were grown continuously without interrupts. Growth was initiated by a 60 second nitridation of the HVPE GaN substrate surface, immediately followed by growth of an ultra-thin, 1.5 nm AlN nucleation layer \cite{Cao3}. Next, a 1 $\mu$m thick beryllium-doped (10$^{19}$ cm$^{-3}$) GaN layer was deposited followed by a 0.3 $\mu$m thick lesser-doped region (2$\times$10$^{17}$ cm$^{-3}$) \cite{Storm},\cite{Deen2}. Next, a 200-nm unintentionally-doped GaN buffer layer was grown. We previously demonstrated that Be compensation-doped layers had minimal deleterious effects on electrical properties when separated from the 2DEG by at least 200 nm \cite{Storm2}. The active heterostructure layers were subsequently grown following an AlN/GaN/AlN sequence with thicknesses of 3/15/3 nm, respectively. These layer thicknesses were chosen in order to avoid lattice relaxation of the strained AlN layers while maintaining an optimal $\mu$-$n_s$ product. It has previously been shown that lattice relaxation begins to occur in the form of micro-cracks along crystalline planes in AlN layers greater than 4 nm when grown pseudomorpically on GaN \cite{Deen1},\cite{Jena1}. Post-growth characterization by atomic force microscopy showed (Fig. \ref{fig:xsection}(c)) a RMS surface roughness of 0.64 nm in a 2$\times$2 $\mu$m$^2$ scan with no indication of lattice relaxation of the AlN layers. An inductive-based contact-less sheet resistance measurement showed as-grown room-temperature sheet resistance across the wafer to be 210 $\Omega/\Box$ indicating 2DEG population in one or both of the channels in the as-grown structure.

\begin{figure}
\includegraphics[width=\columnwidth]{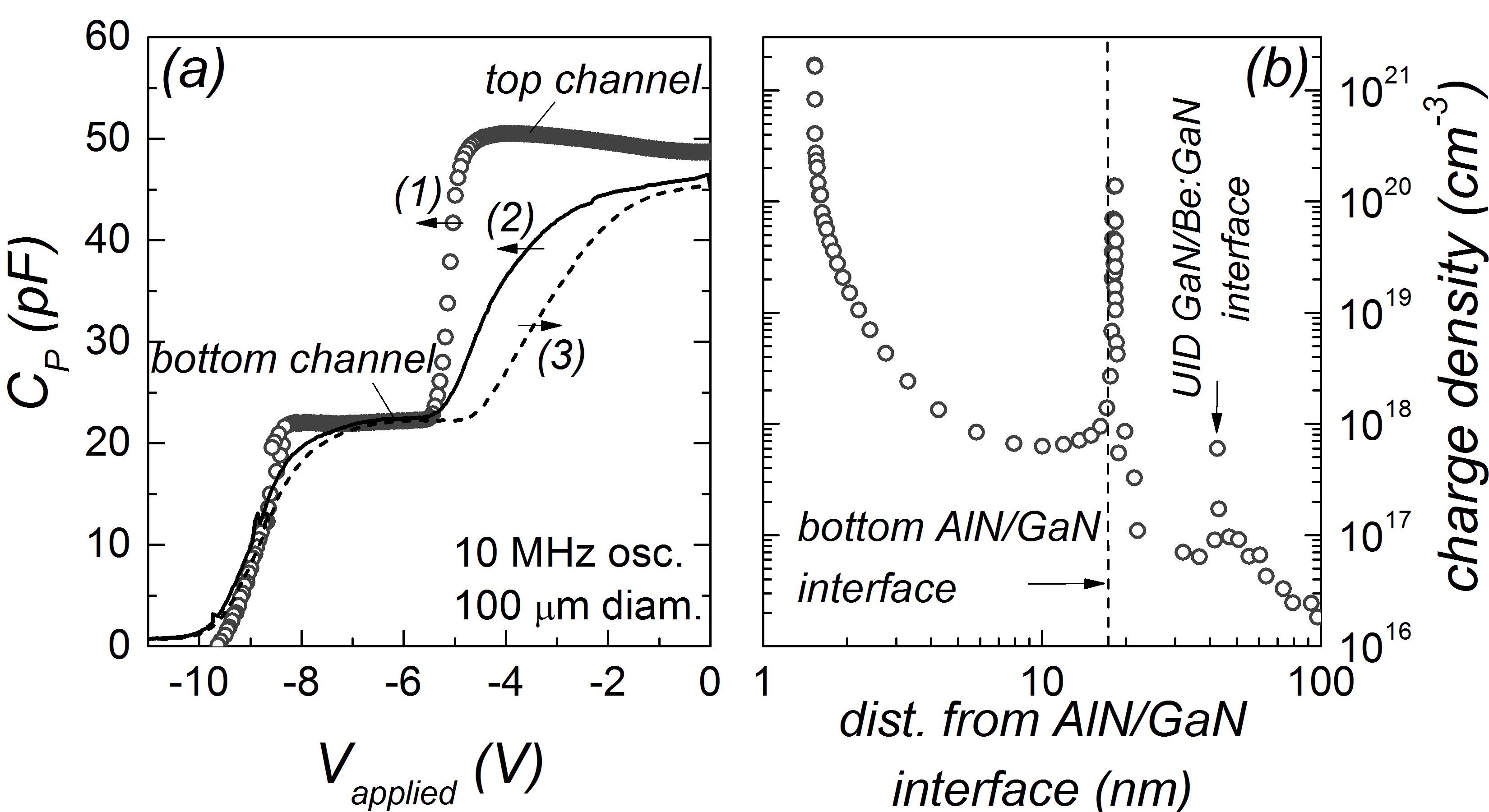}
\caption{(a) Capacitance-voltage characteristics and (b) charge profile for the dual-channel test capacitor showing two distinct capacitance plateaus corresponding to the two parallel 2DEG distributions.}
\label{fig:cv}
\end{figure}
An ohmic-first processing schedule was employed to ensure the best conditions for forming low-resistance ohmic contacts to both parallel 2DEG channels. A pre-metallization Cl-based dry etch was employed to etch through the top AlN and GaN layers prior to contact metallization. The target etch depth was 18 nm below the terminal surface at the interface made between the GaN spacer and the bottom AlN layer. Electron beam evaporation was used to deposit a Ti/Al/Ni/Au metallic layer structure with corresponding thicknesses of 300/20/30/10 nm. An 860$^o$ C rapid thermal anneal was performed for 30 seconds following the metal contact deposition and a contact resistance of $\sim$2 $\Omega$-mm resulted. Mesa and inter-device isolation was facilitated by a Cl-based dry etch. Electron beam lithography was used to define the sub-micron gate-recess on half of the HEMTs on wafer. A Cl-based dry etch was utilized for the gate recess etch. The target depth was 17 nm from the terminal surface such that 1 nm of GaN spacer remained. The gate length was targeted at 300 nm by the recess etch with a 500 nm gate head length. Following the gate recess, a 7 nm thick atomic layer deposited Al$_2$O$_3$ film was used for gate insulation. Optical lithography was also used for the definition of 1 $\mu$m gates and other large-gated test structures following oxide deposition. A Ni/Au gate metal deposition and lift-off concluded the fabrication. Pertinent transistor geometries were source-drain separation ($L_{DS}$) of 3 $\mu$m and gate width ($W_G$) of 150 $\mu$m.

Post-oxidation Hall effect measurements were performed on van der Pauw structures that included both channels and the room temperature sheet resistance and constituent parameters were measured to be $R_{sh}$ =  180 $\Omega/\square$, $n_s$ = 2.3 $\times$ 10$^{13}$ cm$^{-2}$, $\mu$ $\sim$ 1500 cm$^2$/Vs. The measured mobility represents an averaged mobility of both parallel channels since there was no convenient means to differentiate between the channels with the standard on-wafer Hall measurement. However, the individual charge densities of each channel were determined through capacitance-voltage ($CV$) profiling.

$CV$ measurements on a 100 $\mu$m diameter Al$_2$O$_3$/AlN/GaN/AlN/GaN test capacitor at 10 MHz showed two distinct plateaus (Fig. \ref{fig:cv}) indicating two separate charge distributions in the heterostructure. The plot shows three curves with corresponding bias voltage sweep directions. Curve $(1)$ was the initial downward sweep, curve $(2)$ followed in the same direction, and curve $(3)$ was a final upward sweep. A clear deviation is seen between curve $(1)$ and subsequent curves that may amount to charging of interface and bulk traps in the hetero-system \cite{Zhang},\cite{Jena2}. The curve evolution from $(2)$ to $(3)$ shows some hysteresis, which may indicate oxide or confined mobile charge in the GaN spacer layer or buffer layer. It is assumed that curve $(1)$ gives the best evidence of the pristine heterostructure. Therefore, the integration of the lower capacitance plateau in (1) yields a charge density of 1.5 $\times$ 10$^{13}$ cm$^{-2}$ corresponding to the lower 2DEG. The integration of the total $CV$ profile yields a combined charge density of 2.6 $\times$ 10$^{13}$ cm$^{-2}$ which is in agreement with the Hall effect results. The difference gives the upper 2DEG density and was found to be 1.1 $\times$ 10$^{13}$ cm$^{-2}$. $CV$ curve $(1)$ was used to calculate the approximate charge density profile by $n(z) = (C^3/q\epsilon_s)(dC/dV)^{-1}$ and is show in Fig. \ref{fig:cv}(b). These charge densities were used to calibrate the electrostatic conditions used to calculate the band diagrams shown in Fig. 1. A self-consistent Poisson-Schrodinger solver \cite{Snider} was used to calculate band diagrams to show the effect with and without 6$\times$10$^{13}$ cm$^{-2}$ trapped charge at the Al$_2$O$_3$/AlN interface. The trap state density of 6$\times$10$^{13}$ cm$^{-2}$ was chosen for the band diagram simulations based off previous works that have extracted similar trap state densities of the oxide/AlN junction from high-frequency $CV$ methods and have correlated the trap density to spatially-fixed interfacial polarization states of the AlN barrier \cite{Jena2,Deen4,Ibbetson}. The work function ($\Phi_B = \chi_{Ni} - \chi_{ox}$, where $\chi$ is the electron affinity of the designated material layer) used for the Ni-Al$_2$O$_3$ gate was 2 eV and band offsets taken from Ref. \cite{Jena2} were used for the gated heterostructure simulations. Based off the simulation as well as prior work on single-channel AlN/GaN heterostructures \cite{Jena2}, the presence of (ionized positive) donor-like trap states at the oxide/AlN interface cause downward band bending, which promotes an increase in the upper 2DEG density as well as the lower 2DEG density, though with a milder effect. However, it should be noted that though the 2DEG density increases under the influence of ionized surface/interface trap states, the ionized surface states trap electrons from the gate electrode and high-density 2DEG when under bias. Thus, reversing the downward band bending and promoting channel depletion near the gate electrode where the field is the highest. The effect has been referred to as virtual gate extension. The majority of the traps have slow (dis)charge times and cannot respond to gate modulation frequencies in the GHz range or sharp gate pulses. Therefore, in terms of frequency response, they serve to diminish frequency performance of the HEMT through unmodulated channel depletion (at GHz frequency) unless the trap states are either passivated or electrostatically screened. This point motivates the recessed-gate HEMT design reported in this work.

\begin{figure}
\includegraphics[width=\columnwidth]{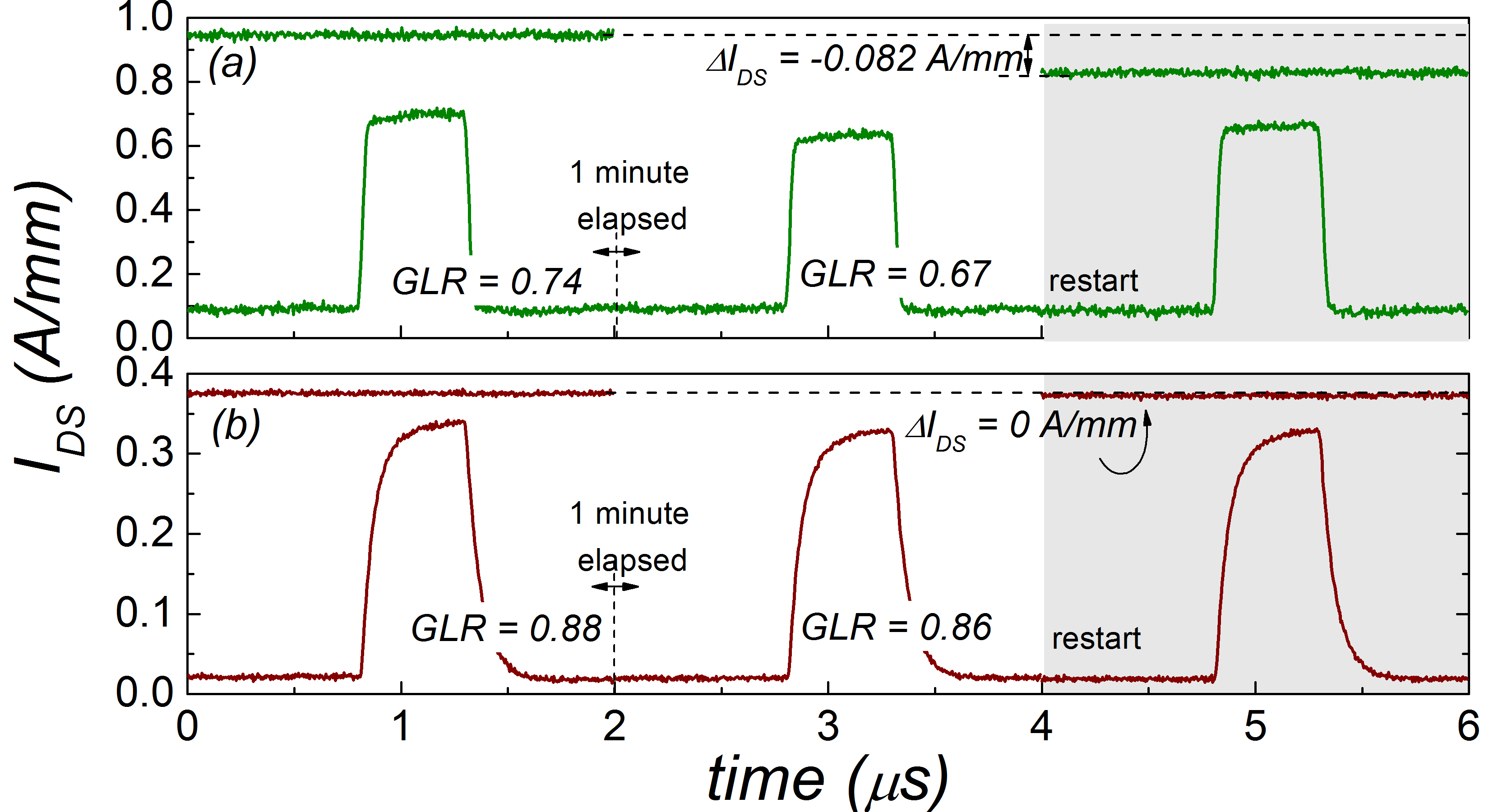}
\caption{Pulsed gate lag characteristics for the (a) dual-channel HEMT with bias conditions of $V_{DS}$ = $10 V$ and $-20V$ $<$ $V_{GS}$ $<$ 0$V$ and (b) the recessed-gate HEMT with bias conditions of $V_{DS}$ = $10 V$ and $-20V < V_{GS} < +3V$. Pulse width for both pulse sequences was 0.5 $\mu$s.}
\label{fig:GLR}
\end{figure}
\begin{figure}
\includegraphics[width=\columnwidth]{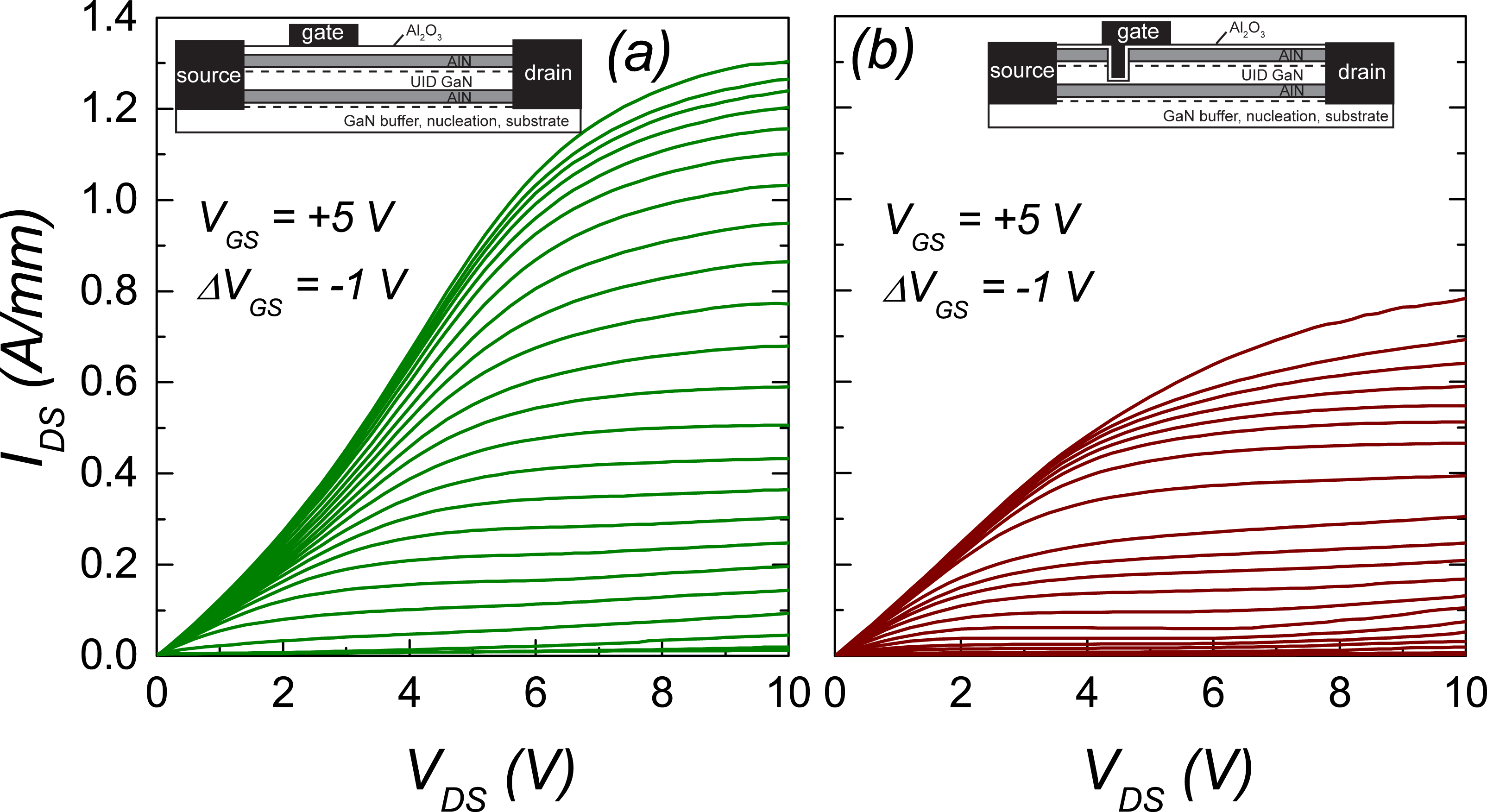}
\caption{Drain characteristics for the (a) dual-channel HEMT and (b) the recessed-gate HEMT.}
\label{fig:drain}
\end{figure}

Gate lag refers to the time delay of a HEMT's drain current recovery in response to a gate voltage pulse. Gate lag results from a slow recovery from depletion of the channel charge due to proximal trapped charge \cite{Binari,Wang}. Interfacial trapped charge such as those at the oxide/AlN interface \cite{Deen4} can lead to gate lag \cite{Binari}. Therefore, a temporally-sequential pulsed gate voltage lag measurement has been used to quantify the gate lag response of the dual-channel and recessed-gate HEMTs. The measurement schedule began by measuring open-channel drain current where predetermined values of $V_{DS}$ and $V_{GS}$ where chosen based off the dc $IV$ characteristics. Those values were $V_{GS} = 0 V$ and $V_{GS} = +3 V$ with $V_{DS} = 10 V$ for the dual-channel and recessed-gate HEMTs, respectively. The measured drain current density in the dc open-channel condition ($I_{DS,o}$) is later used as the normalization value when calculated the gate lag ratio as defined by $GLR = I_{DS,pulse}/I_{DS,o}$. Normalization allows for the comparison of the gate lag response between dissimilar devices. The open-channel drain currents for the stated $V_{GS}$ values used in our measurements were $I_{DS} = 0.95$ A/mm and $I_{DS} = 0.38$ A/mm for the dual-channel and recessed-gate HEMTs, respectively (see Fig. \ref{fig:GLR}). These values were chosen as the drain current resulting at knee voltage under a resistively-loaded drain (see Fig. \ref{fig:drain}). Following the dc $I_{DS,o}$ measurement, $V_{GS}$ is brought to a value within the sub-threshold regime for a prescribed amount of time (0.8 $\mu$s in our schedule), which allows charging of trap states. Then $V_{GS}$ is abruptly pulsed back to the open-channel value previously listed for a specified pulse duration (0.5 $\mu$s in our schedule) and the drain current is monitored during the pulse cycle ($I_{DS,pulse}$) before $V_{GS}$ is brought back into sub-threshold. This concludes the primary gate lag measurement whereby a GLR value may be determined. In our measurement schedule shown in Fig. \ref{fig:GLR}, we have additionally made successive gate pulses 1 minute apart ($V_{GS}$ held in sub-threshold between pulses) in order to observe the effects of higher trapped charge density on GLR. Moreover, our measurement schedule included a restart where all bias voltages where brought to $0 V$ immediately before repeating the measurement schedule just described. This allows for the observation of how degraded the dc $I_{DS,o}$ value has become (grey region in Fig. \ref{fig:GLR}) and serves as a proxy for current slump. The quantity, $\Delta I_{DS} = I_{DS,o} - I_{DS,1}$, where $I_{DS,1}$ is the dc value of $I_{DS}$ measured upon restarting the gate pulse schedule.

The results of pulsed gate lag measurements are shown in Fig. \ref{fig:GLR} for the (a) dual channel AlN/GaN HEMT and the (b) charge-screening HEMT. The dual channel HEMT in (a) demonstrated a GLR of 0.74 which decreased while the HEMT was biased in sub-threshold to 0.67 for subsequent pulses. The hypothesis is that this reduction is mainly due to surface state charging and corresponding depletion of the upper channel (current collapse) since it was shown that the upper channel makes up a large fraction of the total drain current for the dual-channel HEMTs. Upon restarting the measurement the open channel drain current was found to have diminished by 82 mA/mm which is indicative of 2DEG channel depletion and possibly some buffer trapping. The recessed-gate HEMT in (b) demonstrated a GLR of 0.88-0.86, which indicates strong suppression of interface trap related gate lag degradation by its near unity value. Thus, the recessed-gate HEMT showed significant GLR performance improvement over the non-recessed dual-channel HEMT. The traps are assumed to be located at the oxide/AlN interface as is denoted in Fig. \ref{fig:xsection}  \cite{Zhang},\cite{Jena2}. The recessed-gate HEMT showed an emphasized (dis)charge curvature of the drain current pulse in Fig. \ref{fig:GLR}(b). This may be a manifestation of increased gate-to-channel capacitive charging time between the gate metal and upper 2DEG and would corroborate the slower frequency response of the recessed-gate HEMT compared to the dual-channel HEMT in Fig. \ref{fig:freq}(b). Further design enhancements are anticipated to alleviate some of the $RC$ charging in the HEMT design. A notable result of the recessed-gate HEMT is that after the GLR sequence was stopped and restarted, the initial drain current density had not diminished despite the absence of a passivation layer other than the thin Al$_2$O$_3$ gate insulation. Although other reports have been made on nitride-based dual-channel HEMTs with alloyed barrier layers \cite{Chu, Jha, Zhang}, none have included gate lag measurements. To the best of our knowledge, this is the first report on dual-channel AlN/GaN HEMT gate lag as well as its innovation to the recessed-gate HEMT architecture. Although not shown, we have typically observed single channel AlN/GaN HEMTs grown on sapphire or SiC substrates with comparable oxide layer thicknesses to have GLRs of $<$ 0.5. Further refinements in the contacts, gate process, and layer structure are anticipated to advance the design to fully mitigate the detriment of surface traps observed in these ultra-shallow channel AlN/GaN HEMTs.

\begin{figure}
\includegraphics[width=\columnwidth]{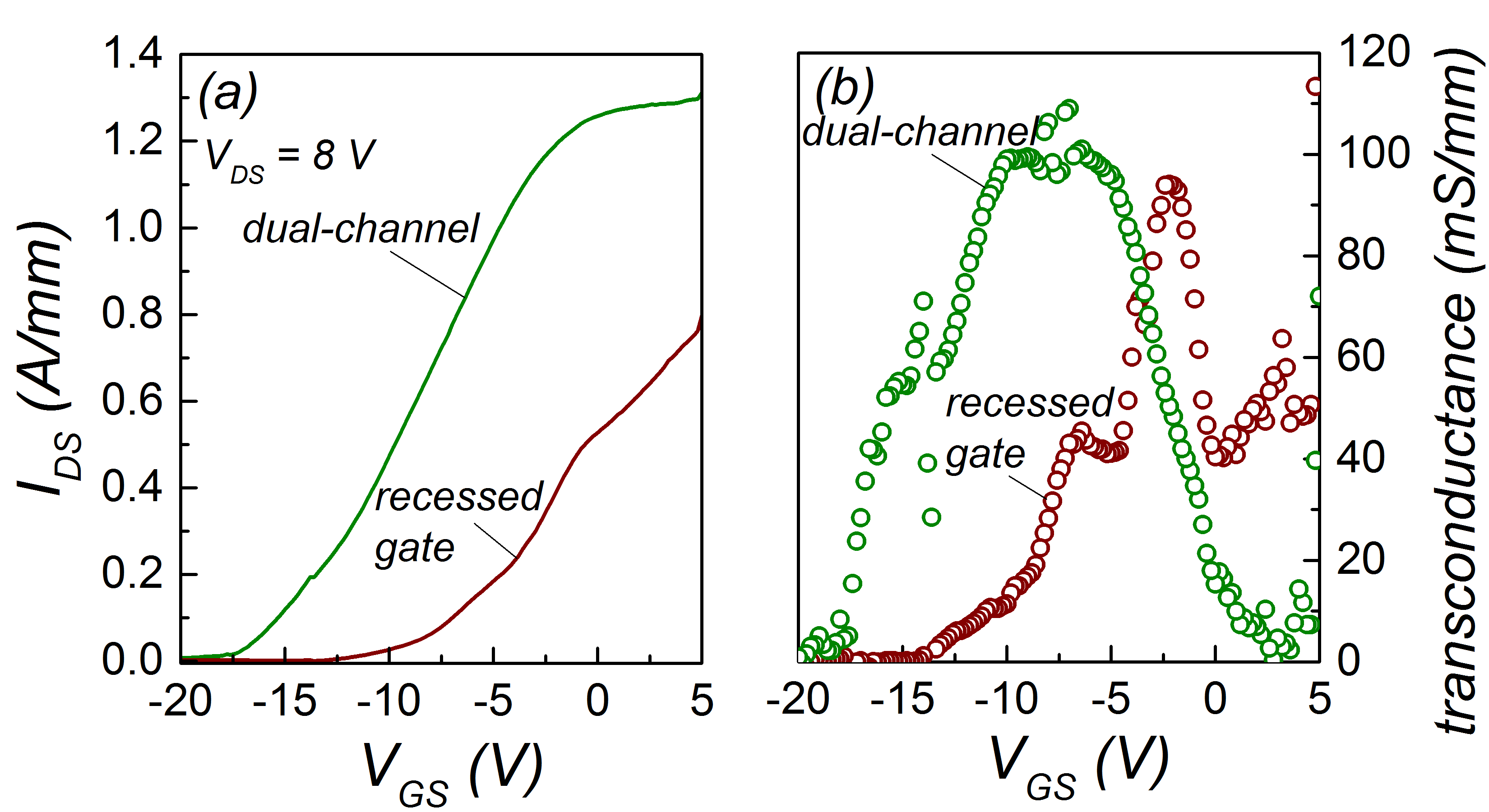}
\caption{(a) Transfer and (b) transconductance characteristics for the dual-channel and recessed-gate HEMTs.}
\label{fig:xfer}
\end{figure}

\begin{figure}
\includegraphics[width=\columnwidth]{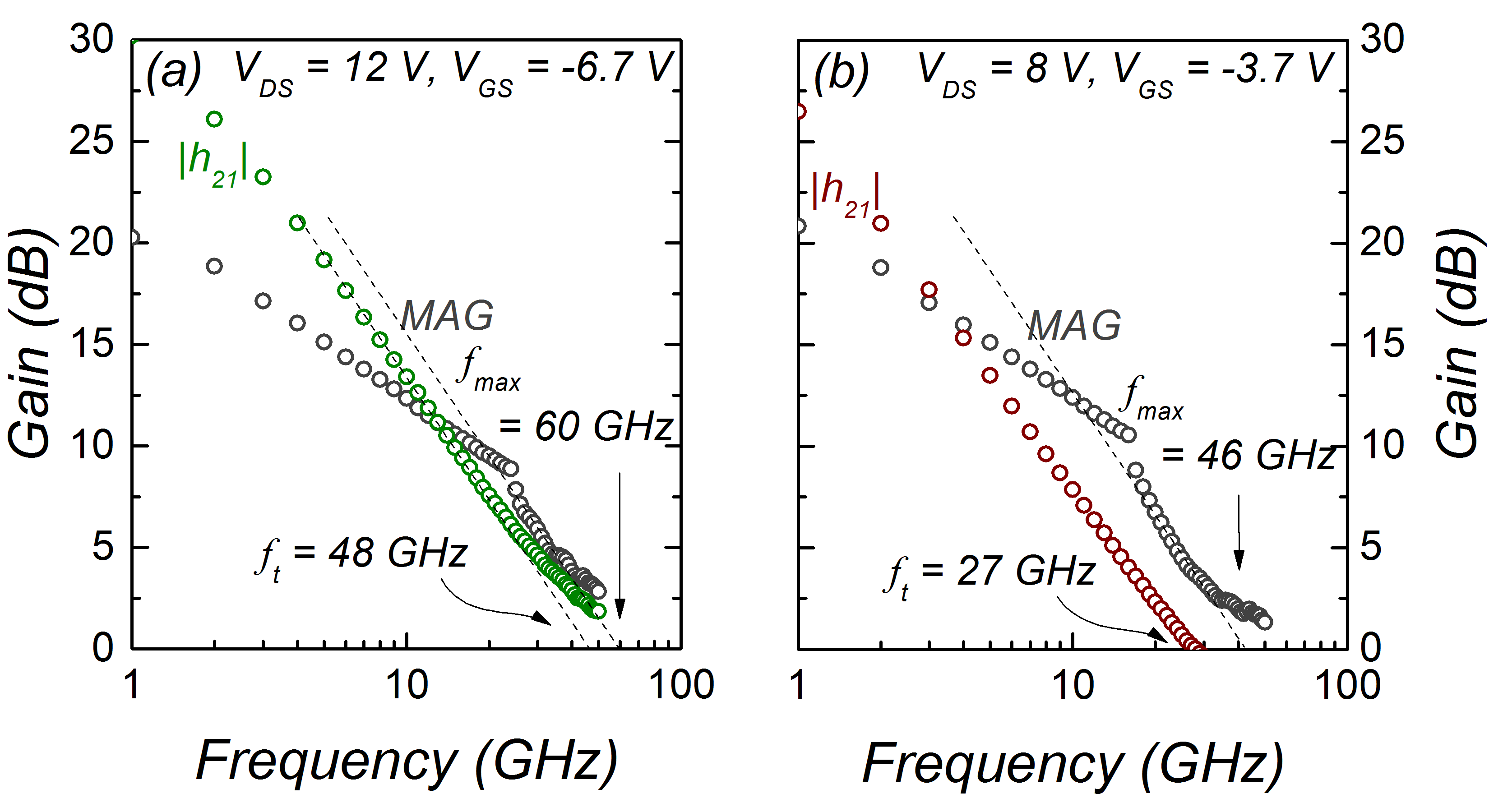}
\caption{Small signal frequency performance of the (a) dual-channel HEMT and (b) recessed-gate HEMT.}
\label{fig:freq}
\end{figure}

Dual-channel and recessed-gate HEMTs were fabricated on the same wafer and standard dc and small-signal rf HEMT characterizations were performed in order to qualify and contextualize the operational performance of the HEMTs through standard techniques. Both HEMT varieties' dc drain characteristics are shown in Fig. \ref{fig:drain} (a) and (b), respectively. The dual channel HEMT had a maximum drain current density of $\sim$2X that of the charge screening HEMT which indicates there was contact made to the etched sidewalls of the top channel. However, slight non-linearity in the drain characteristic at low voltage of the dual-channel HEMT indicates that the contact made was not purely ohmic. The recessed-gate HEMT showed peculiar transfer characteristics with an absence of a well defined maximum current density and non-linear characteristic as shown in Fig. \ref{fig:xfer}(a). Consequently, a multi-peaked $g_m$ was observed as shown in Fig. \ref{fig:xfer}(b) as a result of the non-linear transfer characteristic. The non-linearity in the recessed-gate HEMT transfer characteristic is a manifestation of the two channels in the access region immediately beneath the T-gate head where a small portion of the dual-channel access region (source-side and drain-side) is covered by gate metal (see Fig. \ref{fig:xsection}a). The gate overlap in the access region introduces additional charge from the upper 2DEG that undergoes depletion simultaneously with the conduction channel, a form of capacitive coupling. While capacitive coupling between both 2DEG channels may occur, charge transfer between 2DEGs is not expected for the heterostructure investigated in this work due to the thick GaN spacer and high energy barrier of the buried AlN layer. Capacitive coupling can cause some reduction in frequency performance but can be eliminated by proper engineering of the gate electrode by a self-aligned process that does not have gate overlap in the access region.

Small signal performance was measured up to 50 GHz for both the non-recessed and recessed-gate dual channel HEMTs, as shown in Fig. \ref{fig:freq} (a) and (b), respectively. Unity current gain frequency, $f_t$, and maximum frequency of oscillation, $f_{max}$ were measured to be 27 GHz and 46 GHz, respectively, for the recessed-gate HEMT and 48 GHz and 60 GHz, respectively, for the dual-channel HEMT. The product of $f_t$ and gate length give a measure of average electron velocity in the channel. The resulting $f_t$-$L_g$ products were 8.1 GHz-$\mu$m and 14.4 GHz-$\mu$m for the recessed-gate and non-recessed HEMTs, respectively. The gate length used for the $f_t$-$L_g$ product was 300 nm for both the dual-channel and recessed-gate HEMTs since that length corresponds to the length of the gate stem closest to the channel (Fig. \ref{fig:xsection}(a)). It is noted that the non-recessed dual-channel HEMT demonstrated higher small signal frequency metrics. This is likely due to the recessed-gate HEMT having a larger $C_{GS}$ and $C_{GD}$ resulting in a higher gate charging time. The three charging times involved in setting $f_t$ are the parasitic charging time ($C_{GD}(R_S + R_D$), channel charging time ($(C_{GD}+C_{GS})\times g_{DS}/g_m$), and the drain delay ($C_{GD}/g_m$) \cite{Shinohara,Deen2}. In the case of the recessed-gate HEMT, $C_{GD}$ and $C_{GS}$ increase due to the presence of the upper 2DEG while the lower 2DEG is the only modulated current-carrying channel, and therefore may slightly suffer in terms of frequency performance due to the increased (parasitic) gate capacitance. Nevertheless, the $f_t$-$L_g$ product of 8.1 GHz-$\mu$m is comparable to many reports of GaN-based HEMTs, which implies that if gate capacitance poses an issue, it is not severe. Additionally, a lower access resistance from the two parallel channels may also serve to improve frequency performance for the dual-channel HEMT.

In summary, we have demonstrated a novel recessed-gate dual-channel AlN/GaN/AlN/GaN HEMT architecture that suppresses the deleterious effect surface and interface-originated trapped charges have on drain current recovery under pulsed-gate conditions. This is achieved through the employment of the upper 2DEG as an equipotential that screens the potential fluctuations of the trapped charge. The primary indicator of the recessed-gate HEMT design's efficacy for reducing the effect of surface/interface trapping effects is a GLR of 0.88 demonstrated with no observable decrease in drain current during subsequent gate voltage pulses over time. Small signal performance of 27/46 GHz was achieved for $f_t$/$f_{max}$ in the recessed-gate HEMT with gate lengths of 300 nm. The dual-channel recessed-gate AlN/GaN HEMT demonstrates the feasibility for alternative designs to enhance pulsed-gate performance in HEMTs.

The authors acknowledge N. Green for help with device processing and Professors D. Jena and H. Xing at Cornell University for constructive technical discussion. This work was funded by the Office of Naval Research (P. Maki).

\end{document}